\documentclass[amsmath,amssymb, aps, prl,twocolumn]{revtex4-1}%
\pdfoutput=1
\usepackage{graphicx}
\usepackage{dcolumn}
\usepackage{bm}
\usepackage{float}
\usepackage{epsfig}
\usepackage{amsmath}%
\usepackage{amsfonts}%
\usepackage{amssymb}
\usepackage{natbib}

\begin{document}

\preprint{APS/123-QED}

\title{Maxwell's equations approach to soliton excitations of surface plasmonic resonances}
\author{C. Mili\'{a}n$^{1,2}$, D. E. Ceballos-Herrera$^{3}$, D. V. Skryabin$^{2}$, and A. Ferrando$^{4}$}
\affiliation{
$^{1}$Instituto de Instrumentaci\'on para Imagen Molecular (I3M), InterTech,
Universitat Polit\`ecnica de Val\`encia, Camino de Vera
S/N 46022 Valencia, Spain\\
$^{2}$Centre for Photonics and Photonic Materials, Department of Physics,
University of Bath, Bath BA2 7AY, United Kingdom\\
$^{3}$Universidad Aut\'{o}noma de Nuevo Le\'{o}n, Facultad de 
Ciencias F\'{i}sico Matem\'{a}ticas, Av. Universidad S/N, Cd. Universitaria, Nuevo Le\'{o}n, M\'{e}xico\\
$^{4}$Departament d'\`{O}ptica, Interdisciplinary Modeling Group InterTech,
Universitat de Val\`{e}ncia, Dr. Moliner 50, 46100
Burjassot (Val\`{e}ncia), Spain
}

\begin{abstract}
We demonstrate that soliton-plasmon bound states appear naturally as propagating eigenmodes of nonlinear Maxwell's equations for a metal/dielectric/Kerr interface. By means of a variational method, we give an explicit and simplified  expression for the full-vector nonlinear operator of the system. Soliplasmon states (propagating surface soliton-plasmon modes) can be then analytically calculated as eigenmodes of this non-selfadjoint operator. The theoretical treatment of the system predicts the key features of the stationary solutions and gives physical insight to understand the inherent stability and dynamics observed by means of finite element numerical modeling of the time independent nonlinear Maxwell equations. Our results contribute with a new theory for the development of power-tunable photonic nanocircuits based on nonlinear plasmonic waveguides.
\end{abstract}

\pacs{Valid PACS appear here}
\maketitle

Nonlinear plasmonics brings the richness of the nonlinear dynamics into the \textit{nano} scaled world of photonics. The interest in finding nano scaled optical solitons has been prominent in the last few years. Recent studies have demonstrated
self compensated diffraction beams on a single metal/dielectric (MD) interface \cite{Davoyan_OE_2009}, plasmon-solitons in MDM configurations \cite{Feigenbaum_OL_2007}, solitons in systems with gain and loss \cite{Marinis}, in waveguide arrays \cite{Ye_PRL_2010,Salgueiro_APL_2010,Milian_APL_2011}, and in chains of nano particles \cite{Noskov_PRL_2012} and Hydrogen atoms \cite{DePrince_PRL_2011}. Typically, the nonlinear response is directly associated to the strong field of the SPP close to the interface, giving rise to surface enhancement effects \cite{Skryabin_JOSAB_2011}.

\begin{figure}
\includegraphics[scale=0.25]{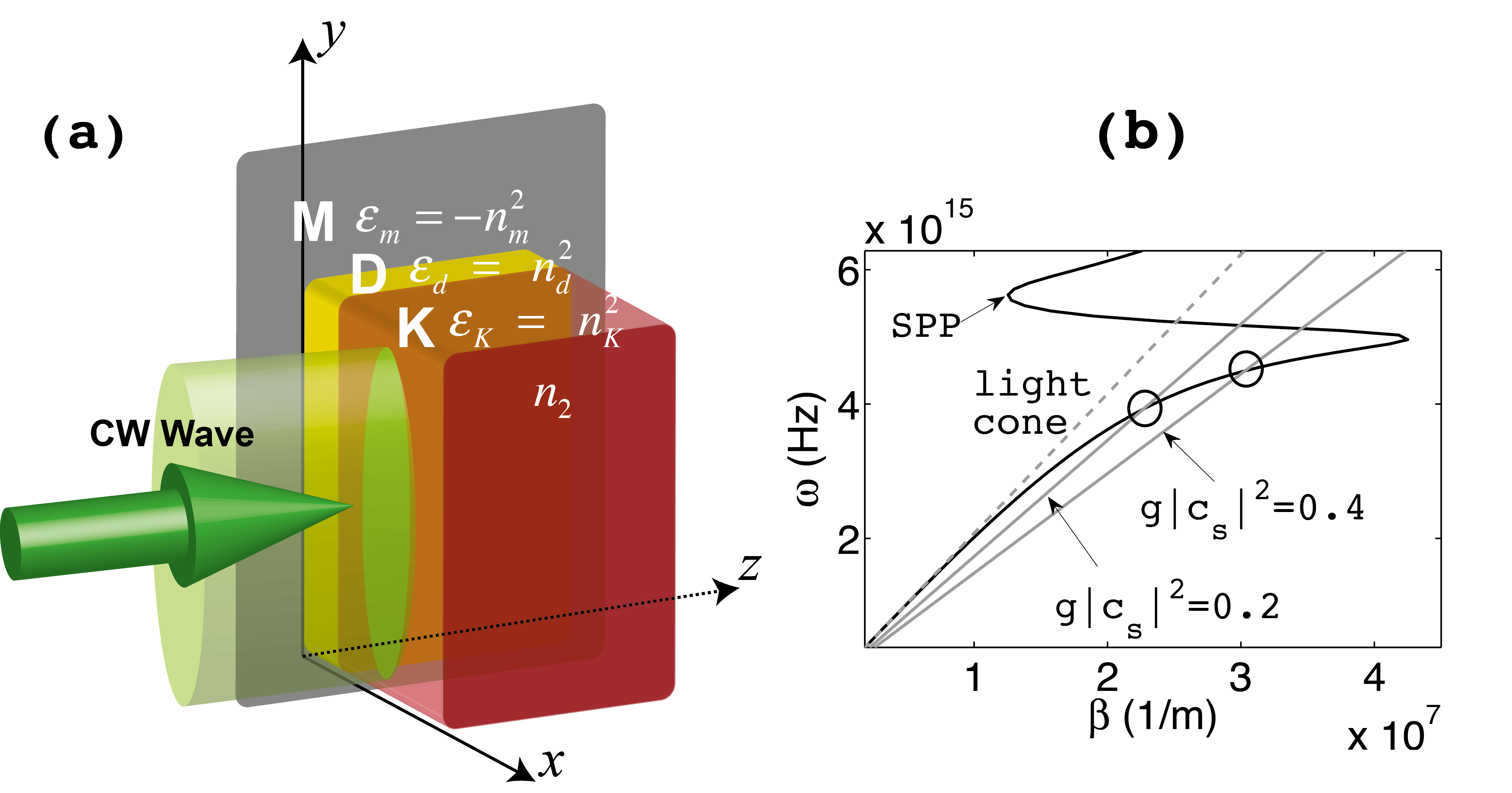}
\caption{(color online). (a) metal/dielectric/Kerr (MDK) structure with
linear dielectric constants $\varepsilon_{m}=-n_{m}^{2}$, $\varepsilon_{d}=n_{d}^{2}$
and $\varepsilon_{K}=n_{K}^{2}$, respectively. (b) Dispersion of a SPP in a lossy and dispersive metal (black) and a spatial
soliton (gray) owning two different amplitudes. Circles enclose the matching
points $(\beta_m,\omega_m)$ and the dashed line marks the light cone $\omega=\beta c/\sqrt{\varepsilon_d}$.}
\label{f1} 
\end{figure}

SPP waves represent one of the most confined form of light known nowadays and therefore conceived as the signals for nano-scaled circuits \cite{Bozhevolnyi_NAT_2006}. Their sub-wavelength confinement is related to the fact that they are coupled light-plasma waves with a propagation constant greater than that of the light cone
\cite{Maier_book_2007} and require, in general, complex setups for their excitation via linear
\cite{Zayats_PR_2005} or nonlinear \cite{FWM_SPP_PRL} mechanisms. In the context of SPP excitation, we demonstrate that nonlinearities play a subtle role because, similarly to SPP's, spatial solitons in bulk dielectric media also have a propagation constant above that of the plane waves at the same frequency \cite{Kivshar_book_2006}. Hence, these two waves can couple \cite{Bliokh_PRA_2009}, given that their dispersion relations intersect (see Fig. \ref{f1}b), so, even in the absence of strong SPP fields, a nonlinear resonant excitation of a SPP can occur via a monochromatic beam. In order to show explicitly the last statement, and without loss of generality, most of the SPP field is located in a linear dielectric layer in our analysis.

In this letter, we analyze from first principles the properties of the stationary states of nonlinear Maxwell's equations for a metal/dielectric (linear)/Kerr (MDK) interface (see Fig. \ref{f1}a). We use a variational ansatz, which allows us to diagonalize the full vector nonlinear operator of the system and derive a simplified expression for it. The eigenmodes of this operator, the \textit{soliplasmons}, can be understood as stationary nonlinear hybrid states formed by the coupling between a SPP and a spatial soliton (which are not orthogonal). The non-selfadjoint character, in general, of this operator, which is an inherent property of the vectorial nature of the system  (due to the presence of the SPP field), has important physical implications being responsible of an asymmetric coupling between the plasmon and the soliton. This essential feature was not captured by previous heuristic models \cite{Bliokh_PRA_2009} introduced by analogy with coupled linear SPP's \cite{Bliokh_RMP_2008} further analyzed in the context of Josephson junctions \cite{Eks_PRA_2011}.

The simplified model presented here brings the relevant physics associated to the soliplasmon solutions, and enables the
simulation of realistic devices. In particular, it shows (i) that the soliton tail at the metal interface exerts the
main driving action of the coupling, (ii) anti-crossing in the dispersion relations,
(iii) a soliplasmon resonant amplitude such that it becomes a nonlinear SPP wave as a
particular solution, and (iv) a neat classification of the solutions in terms of the relative phase between the SPP
and soliton components. Moreover, first principle modeling, based on finite element analysis, is used to integrate the time
independent vectorial and nonlinear Maxwell equations in two dimensions. Simulations are used to analyze stability of soliplasmons and to show the associated dynamics, which can be understood in terms of the variational model. Although derived here for the monochromatic case, its extension to describe plasmonic pulses \cite{PULSED_SPP} will bring insight to further study soliton dynamics and frequency conversion effects involving these hybrid plasmon-soliton waves.

We analyze the full-vector nonlinear equation for a monochromatic wave (\emph{cw}) with frequency $\omega=ck$ described by its
electric field $\mathbf{E}_{\omega}$:

\begin{eqnarray}
\left[\frac{\partial^{2}}{\partial z^{2}}+\frac{\partial^{2}}{\partial x^{2}}+k^{2}\varepsilon_{L}(x)\right]\mathbf{E}_{\omega} & =\mathcal{L}_{v}(x)\mathbf{E}_{\omega}+ & \mathbf{P}_{\mathrm{NL}}\label{eq:full_vector_NL_eq}
\end{eqnarray}

where $\mathcal{L}_{v} \equiv\nabla\left[\nabla\circ\right]$ is the vector linear operator and
$\mathbf{P}_{\mathrm{NL}}=-k^{2}\chi^{(3)}/3\left\{
2|\mathbf{E}_{\omega}|^{2}\mathbf{E}_{\omega}+\mathbf{E}_{\omega}^{2}\mathbf{E}_{\omega}^{*}\right\}$. The MDK geometry is illuminated from the Kerr medium with a beam parallel to the interface (see Fig. \ref{f1}a) and diffraction along $y$ is neglected. Below, we find analytical solutions for this problem by means of dynamical equations derived from
Eq.\eqref{eq:full_vector_NL_eq} using a variational method. The choice of the variational \emph{ansatz} is motivated by the heuristic model in Ref.\cite{Bliokh_PRA_2009}, according to which soliton-SPP states are expected to exist, and thus taken in the form:

\begin{equation}
\mathbf{E}_{\omega}(x,z)=\left[c_{p}(z)\mathbf{e}_{p}(x)+\hat{\mathbf{u}}c_{s}(z)f_{s}\left(x-a;c_{s}(z)\right)\right]e^{ikn_{K}z}.\label{eq:ansatz}
\end{equation}

\begin{figure}
\includegraphics[scale=0.25]{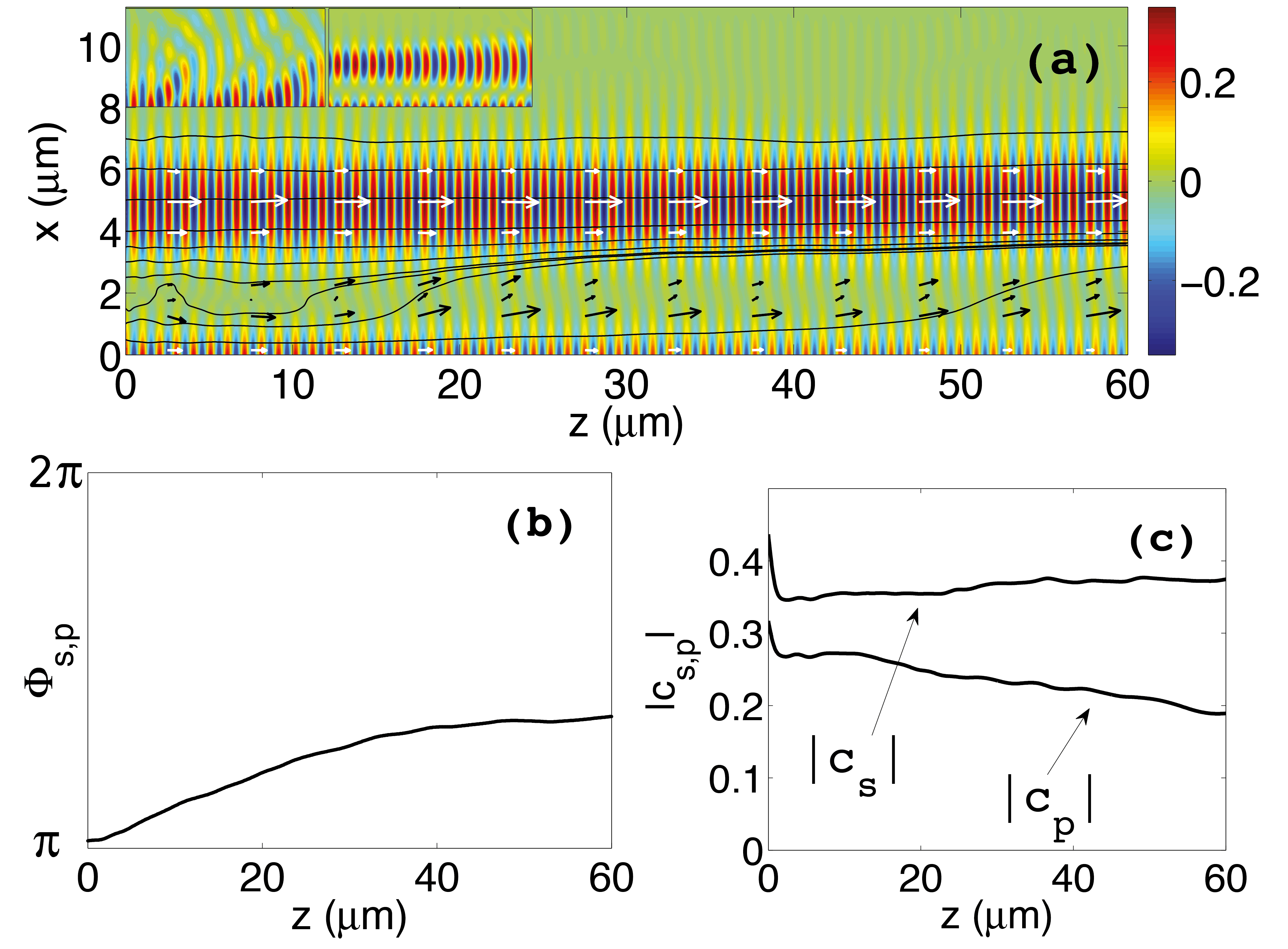}
\caption{(color online). (a) $|\mathcal{E}_{x}|$ of the $n_{\mathrm{eff}}=1.464$ soliplasmon  propagating along $60\
\mu$m. The inset on the left magnifies $|\mathcal{E}_{x}|$ over the area where it
is placed, showing the noise is ejected away. Stream lines show the power
flow, which magnitude is represented by the arrows. Black arrows are
magnified with respect to the white ones. Inset on the right shows
the diffraction observed in the linear propagation over the first $12\mu$m. (b) Soliton-plasmon phase, and (c) soliton and plasmon amplitudes. Sharp initial jumps are due to the large input noise.}
\label{f4} 
\end{figure}

\begin{figure}
\includegraphics[scale=0.25]{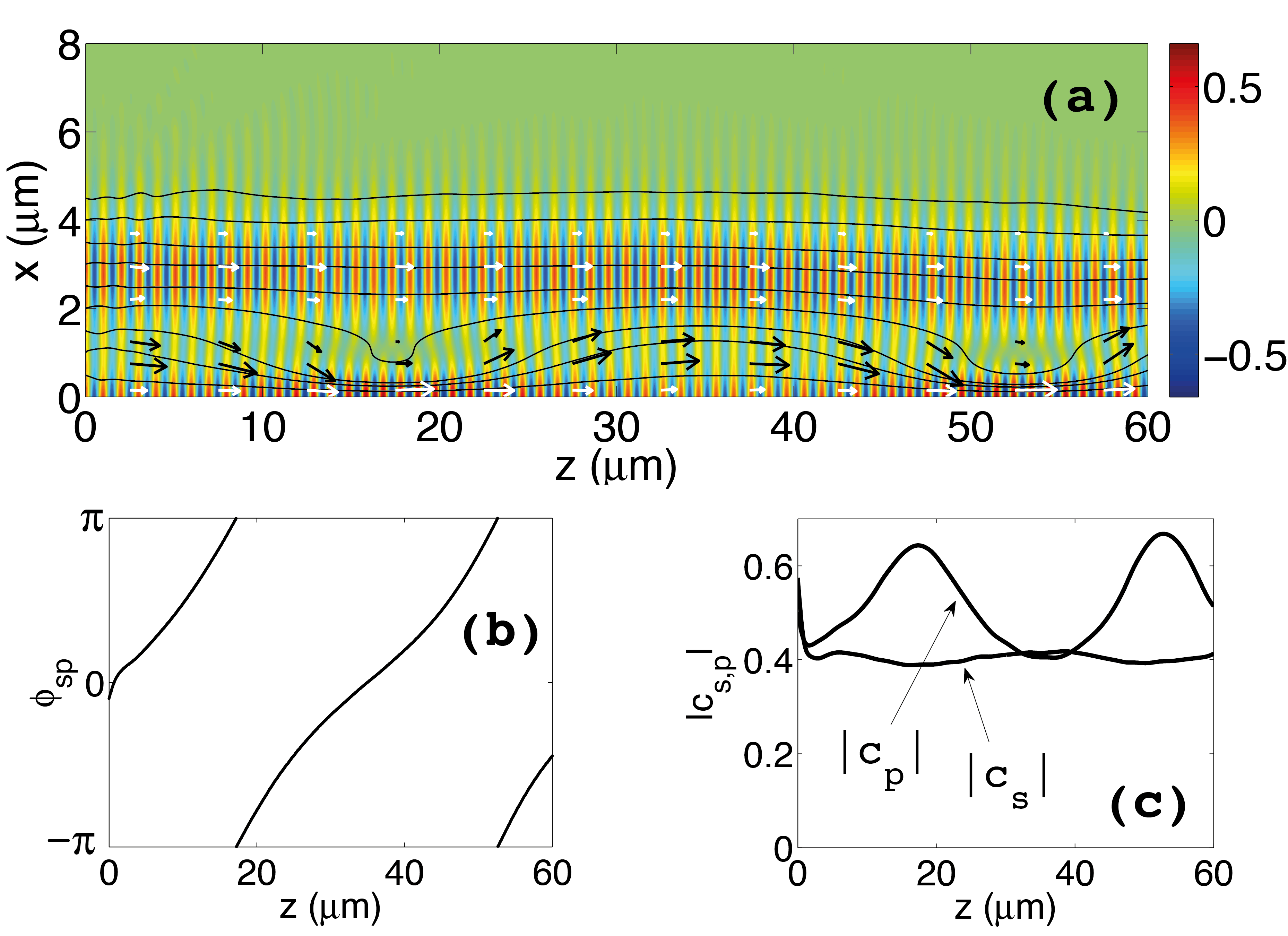}
\caption{(color online). $60\ \mu$m propagation of the $n_{\mathrm{eff}}=1.472$ 0-soliplasmon.
(a) Electric field norm $|\mathcal{E}_{x}|^{2}$, (b) soliton-plasmon relative phase, (c) soliton and plasmon amplitudes.}
\label{f5}
\end{figure}

\begin{figure}
\includegraphics[scale=0.25]{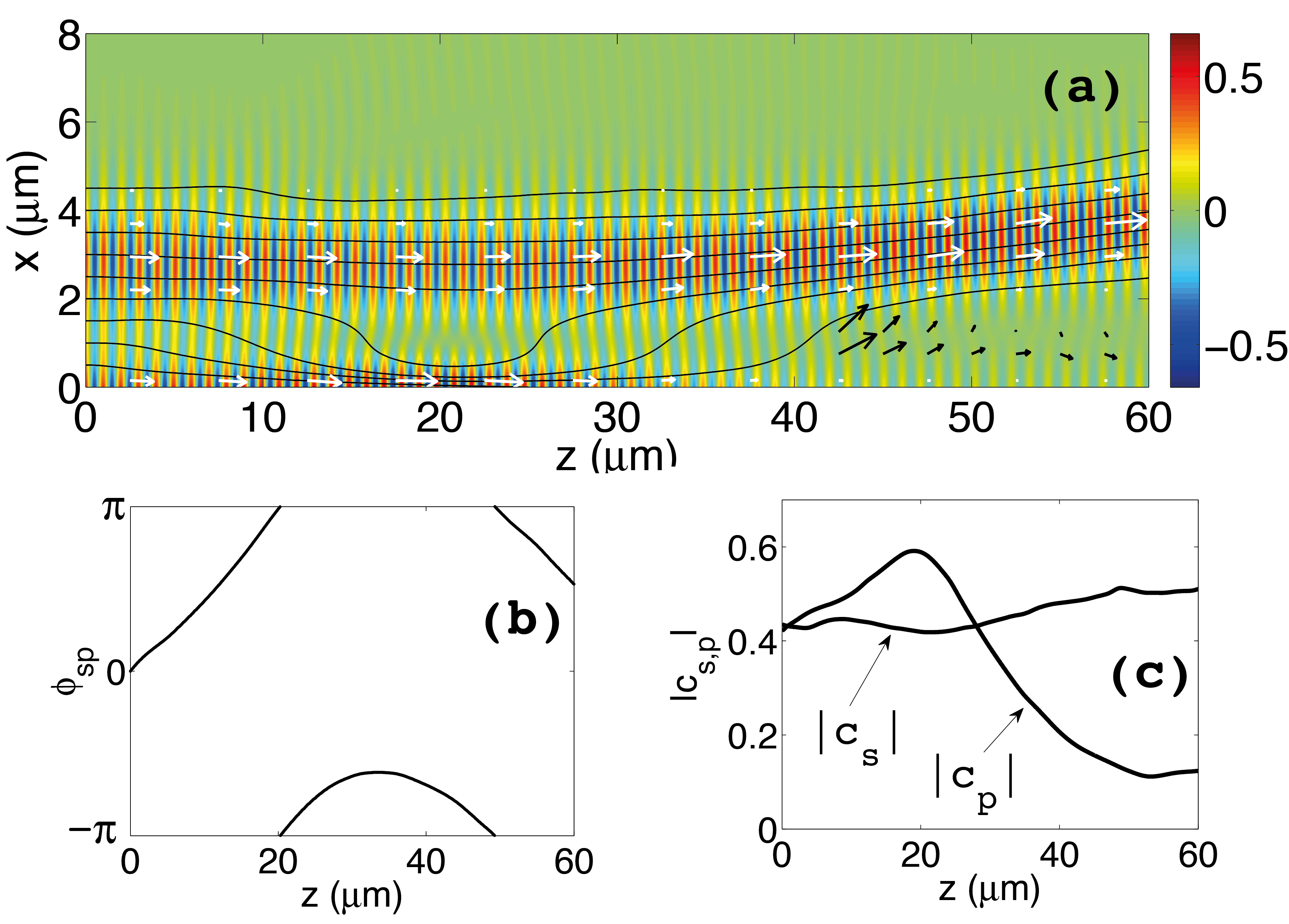}
\caption{(color online). Soliplasmon propagation under the initial conditions of Fig.
\ref{f5},
but accounting for the metal losses, $\epsilon_{m}=-82+i8.3$.}
\label{f6} 
\end{figure}

The field $\mathbf{e}_{p}(x)=[e_{px}(x),e_{pz}(x)]$ corresponds
to a TM-SPP on the MD interface with propagation constant
$\beta_{p}=k\sqrt{\varepsilon_{m}\varepsilon_{d}/\left[\varepsilon_{m}+\varepsilon_{d}\right]}$, which is a stationary
solution of Eq.\eqref{eq:full_vector_NL_eq} with $\mathbf{P}_{\mathrm{NL}}=\mathbf{0}$,
and hence the complex amplitude $c_{p}(z)=c_{p}(0)e^{i\mu_{p}z}$ ($\beta_{p}=kn_{K}+\mu_{p}$). The dielectric function
$\varepsilon_{L}(x)$
is in this case that of the MD interface. The soliton term in Eq.\eqref{eq:ansatz},
$\phi_{s}=c_{s}f_{s}=c_{s}\mathrm{sech}\left(\sqrt{kn_{K}\gamma}\left|c_{s}\right|\left[x-a\right]\right)$,
is located at a distance $a$ from the MD interface and it is a solution of the stationary \emph{scalar} ($\mathcal{L}_{v}=0$) and paraxial 
nonlinear Schr\"{o}dinger equation with $c_{s}(z)=c_{s}(0)e^{i\mu_{s}z}$:
$\left\{1/[2kn_{K}]\partial_{x}^{2}+\gamma\left|\phi_{s}\right|^{2}\right\}\phi_{s}=\mu_{s}\phi_{s}$, so
its wave-number is $\beta_{s}=kn_{K}+\mu_{s}$, with $\mu_{s}=\gamma\left|c_{s}(0)\right|^{2}/2$,
$\gamma=kn_{K}\chi^{(3)}/2=k\varepsilon_{0}cn_{2}n_{K}/2$. It is considered that the overlapping of the SPP and soliton fields
is always small (weak coupling) and the SPP behaves linearly in the
absence of coupling. The weak coupling is automatically satisfied when the dielectric layer is
present, although it can be also achieved in MK structures if $a$ is big enough. Full-vector numerical simulations show
that in this situation
the field is mainly transverse in the nonlinear region. Thus, to a
first approximation, $\mathbf{u}\approx(1,0)$. Our \emph{ansatz} has
only two variational parameters: $c_{p}(z)$ and $c_{s}(z)$. However,
$c_{s}(z)$ appears nonlinearly in Eq. \eqref{eq:ansatz}, preventing
the soliplasmon from behaving as a linear superposition of
modes.  Parallel illumination of the MD interface ensures paraxiality unless high energy transfer takes place. For
weakly coupled quasi-stationary soliplasmon states this is true, justifying the
paraxial approximation and the fact that the soliton position, $a$, is not considered as a variational parameter. These two approximations are supported by our full-vector modeling (see Figs. \ref{f4}-\ref{f6}). 

We substitute the \emph{ansatz} Eq.\eqref{eq:ansatz} in the paraxial
version of Eq.\eqref{eq:full_vector_NL_eq} ($\partial_{z}^{2}\rightarrow2ikn_{K}\partial_{z}-k{}^{2}n_{K}^{2}$) for
$e_{x}\equiv\hat{\mathbf{x}}\mathbf{E}_{\omega}(x,z)$, $-i\partial_ze_x=\hat{M} e_x$, and distinguish between two
regions: linear ($x\le d$) and nonlinear ($x>d$), $d$ being the dielectric layer width. For $x\le d$, $\mathbf{P}_{\mathrm{NL}}=\mathbf{0}$ and we take into account:
(i) $\mathbf{e}_{p}$ is an eigenfunction of the linear
operator with $\mathcal{L}_{v} = -\nabla\left[\varepsilon_{L}^{-1}\nabla\varepsilon_{L}\circ\right]$ and eigenvalue $\beta_{p}^{2}$;
(ii) $\phi_{s}$ is an eigenfunction of the scalar nonlinear operator (vector effects are negligible for $\phi_s$) with eigenvalue $\mu_{s}$,
but behaves linearly in this region ($a>d$). We then project the resulting equation into the plasmon
component by multiplying it by $e_{px}$ and integrating over $x\in\left]-\infty,\infty\right[$. Since the soliton and SPP modes
are not orthogonal ($\int_xe_{px}^{*}f_s\neq0$) we obtain a coupled equation for $dc_{p}/dz$
and $dc_{s}/dz$. An analogous procedure applied in the region $x>d$,
projecting now with respect to $f_{s}$, gives a second equation for $dc_{p}/dz$, $dc_{s}/dz$. Linear combinations of both
equations in the weak coupling approximation give the variational equations of the soliplasmon system ($\left|c\right\rangle \equiv[c_{p},c_{s}]^{T}$):

\begin{equation}
-i\frac{d}{dz}\left|c\right\rangle = M \left|c\right\rangle ,\ \ M =\left[\hat{M}\right]= \left[\begin{array}{cc}
\mu_{p} & q(|c_{s}|)\\
\bar{q}(|c_{s}|) & \mu_{s}
\end{array}\right],\label{eq:variational_eqs}
\end{equation}

where $\mu_{p}\approx\beta_{p}-kn_{K}$ and $\mu_{s}(|c_{s}|)\equiv\gamma|c_{s}|^{2}/2$
($\beta_{s}\approx kn_{K}+\mu_{s}\equiv kn_{K}[1+g|c_s|^2]$). Note the dynamical propagation constants are approximately the
stationary ones due to the weak coupling assumption. We emphasize that the origin of the off diagonal terms (coupling) in Eq. \eqref{eq:variational_eqs} is the non-orthogonality between the plasmon and soliton. Unlike in Ref.\cite{Bliokh_PRA_2009}, the coupling is not symmetric
in general ($M_{12}\neq M_{21}$) because $\hat{M}$ is not a hermitian operator, in fact $\bar{q}=qN_{p}/N_{s}$, where $q\equiv\ k/[2n_{K}N_{p}]\int_{x}\left[\varepsilon_{L}-n_{K}^{2}\right]e_{px}f_{s}$,
$N_{p}\equiv\int_{x}|e_{px}|^{2}$ and $N_{s}\equiv\int_{x}f_{s}^{2}$. Eqs. \eqref{eq:variational_eqs} have
no free parameters, all variational constants are given in terms of
$\varepsilon_{m}$, $\varepsilon_{d}$, $\varepsilon_{K}$ and $n_{2}$:  $N_{p}=\left[\varepsilon_{d}^{2}\kappa_{d}+\varepsilon_{m}^{2}\kappa_{m}\right]/\left[2\varepsilon_{d}^{2}\varepsilon_{m}\kappa_{d}\kappa_{m}+2\varepsilon_{d}\varepsilon_{m}^{2}\kappa_{d}\kappa_{m}\right]$,
where $\kappa_{m,d}=\sqrt{\beta_{p}^{2}-k^{2}\varepsilon_{m,d}}$, and $N_{s}=2/\kappa_{s}$, where $\kappa_{s}\equiv\left[kn_{K}\gamma\right]^{1/2}|c_{s}|$. Remarkably, the soliplasmon coupling is nonlinear in $\left|c_{s}\right|$

\begin{equation}
q\left(\left|c_{s}\right|\right)\approx\frac{\beta_{p}}{N_{p}n_{K}}\frac{\varepsilon_{m}-\varepsilon_{K}}{\kappa_{m}+\sqrt{kn_{K}\gamma}|c_{s}|}\exp\left(-\sqrt{kn_{K}\gamma}a|c_{s}|\right),
\end{equation}

and is driven by the exponentially decaying soliton tail. As a consequence, we predict that a strong soliton in an MDK system will excite a weak SPP ($N_s\gg N_p$, $|c_s|\gg|c_p|$) at a rate $q$, whereas a strong SPP will excite a weak soliton at the much smaller rate $\bar{q}\sim|c_s|$.

Soliplasmons are stationary states of Eq.\eqref{eq:variational_eqs}
given by the eigenvalues ($\mu$) and eigenvectors ($\left|\mu\right\rangle $)
of the matrix $M$. The wavenumber $\beta=kn_{K}+\mu$ is determined from

\begin{equation}
\mu_{\delta}=\bar{\mu}+e^{i\delta}\sqrt{\Delta_{\mu}^{2}+q\bar{q}},\ \delta=0,\pi\label{eq:eigen_mu},
\end{equation}

where $\bar{\mu}\equiv[\mu_{p}+\mu_{s}]/2$ is the propagation constant
mean value and $\Delta_{\mu}\equiv[\mu_{p}-\mu_{s}]/2$ the soliton-plasmon
wave-number mismatch. For an arbitrary $c_{s0}$, the associated eigenvectors are $\left|\mu_{\delta}\right\rangle =c_{s0}\left[q/\left\{\mu_{\delta}-\mu_{p}\right\},\ 1\right]^{T}$ and so

\begin{equation}
\frac{E_{x}(x,z)}{c_{s0}}=\left\{ \frac{q}{\mu_{\delta}-\mu_{p}}e_{px}(x)+\mathrm{sech}(\kappa_{s}[x-a])\right\}
e^{i[kn_{k}+\mu_{\delta}]z}\label{eq:spEx}.
\end{equation}

\begin{figure}
\includegraphics[scale=0.25]{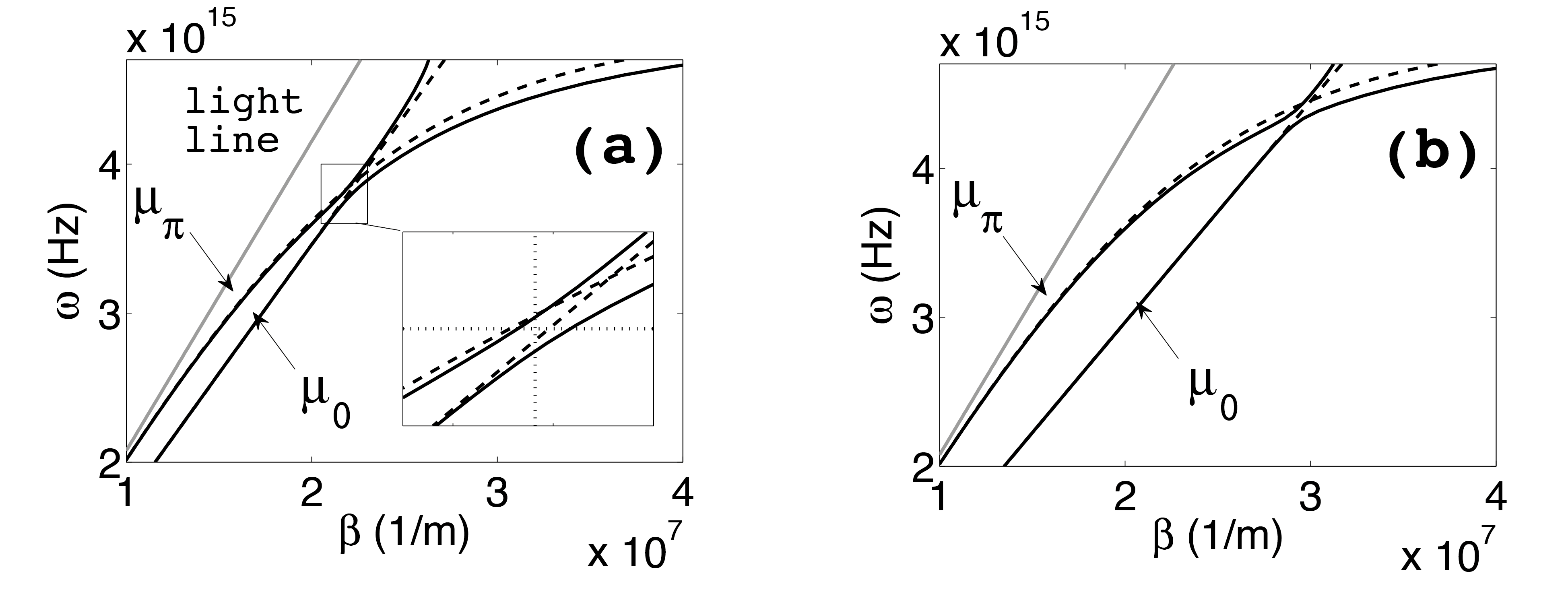}
\caption{Soliplasmon dispersion for soliton peak power levels (a) $g|c_{s}|^{2}=0.2$
and (b) $g|c_{s}|^{2}=0.4$. Dashed lines correspond to dispersion
of soliton and SPP, separately (as in Fig. \ref{f1}b). Inset in (a)
zooms in the anti-crossing region and the intersection of the dotted
lines marks the predicted anti-crossing point. Soliton distance is fixed to $a=3/k_d$ in both figures.}
\label{f2} 
\end{figure}

Note from Eq. \eqref{eq:eigen_mu} that $\mu_{0}>\mu_{p}$ and $\mu_{\pi}<\mu_{p}$, so
the plasmon term in Eq. \eqref{eq:spEx} is positive (negative) for $\delta=0$ ($\pi$). $\delta$
has the meaning of the relative phase between the plasmon and soliton components. Equation \eqref{eq:eigen_mu} provides the dispersion relations in the implicit form
$\omega$ vs $\beta_{\delta}=n_K\omega/c+\mu_{\delta}\left(\omega;\left|c_{s0}\right|,a\right)$, which clearly present the anti-crossing behavior of the $\delta=0,\pi$ branches (see Fig.\ref{f2}). Far from the matching point $\{\beta_m,\omega_m\}$ (see Fig. \ref{f1}b), these curves tend to the non-interacting soliton and plasmon (dotted curves) and the $\delta=0$ ($\pi$) soliplasmons switch from soliton (plasmon) to plasmon (soliton) behavior as $\omega$ increases through $\omega_m$, which is tuned with the anti-crossing point by $|c_{s0}|$.

\begin{figure}
\includegraphics[scale=0.4]{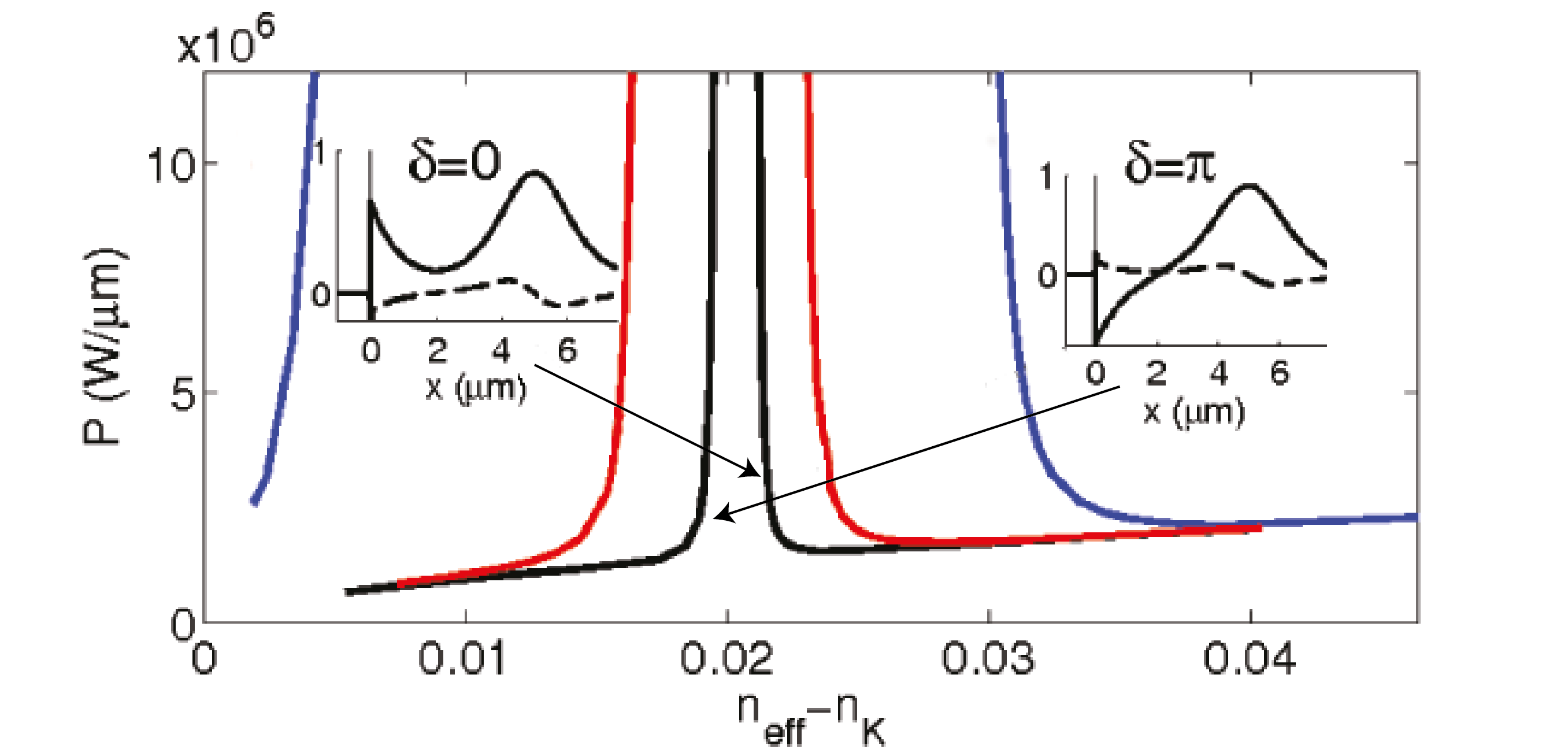}
\caption{(color online). $P$ vs $\mu/k$ curves for $\delta=0$ and $\delta=\pi$ soliplasmon
families at $\omega\simeq1.26$ PHz ($\lambda=1.5\,\mu\mathrm{m}$) for: $a=3\,\mu\mathrm{m}$ (blue),
$a=4\,\mu\mathrm{m}$ (red), and $a=5\,\mu m$ (black). Insets: $x$ (solid) and $z$ (dashed) dimensionless components of typical
mode profiles $\mathcal{E}\equiv\sqrt{\chi^{(3)}}\mathbf{E}$ for $\delta=0$ (right) and $\delta=\pi$ (left).
}\label{f3}
\end{figure}

The soliplasmon guided power is
$P=\int_{x}E_{x}^{*}H_{y}/2=\omega\varepsilon_{0}/[2\beta]\int_{x}\varepsilon\left|E_{x}\right|^{2}$,
where $\varepsilon\equiv\varepsilon_L+\varepsilon_{NL}$.  $P$ vs $\mu$ curves for $0$, $\pi$ soliplasmons can be explicitly calculated using the variational
solutions in Eqs.\eqref{eq:eigen_mu} and \eqref{eq:spEx}: 

\begin{equation}
P_{\delta}\left(\mu\right)\approx P_{\mathrm{s}}+\frac{c\varepsilon_{0}\varepsilon_{d}}{2n_{K}}\left|c_{s0}\right|^{2}\left\{\left[\frac{q}{\mu_{\delta}-\mu_{p}}\right]^{2}N_{p}+\frac{2qq_{ps}}{\mu_{\delta}-\mu_{p}}\right\},\label{eq:P_vs_mu}
\end{equation}

where $P_{\mathrm{s}}$ is the soliton power, $q_{ps}\equiv\int_{x}e_{px}f_{s}$,
and higher order overlapping terms are neglected. The $P(\mu;\omega,a)$ curves (see Fig. \ref{f3}) have the two branches $\mu_{0}>\mu_{p}$ (right) and $\mu_{\pi}<\mu_{p}$ (left) separated by the vertical line $\mu=\mu_{p}$. Far from this asymptote 
($\left|\mu_{\delta}-\mu_{p}\right|/k\gg1$) both branches coalesce into the monotonically increasing soliton curve $P_{s}(\mu)\sim\mu^{1/2}$. However, near the resonance $\mu\approx\mu_{p}$ both curves present divergent slopes, since the singular SPP term in Eq.\eqref{eq:P_vs_mu} dominates, and a gap in $\mu$ is created. According to Eq.\eqref{eq:eigen_mu}, this is proportional to $\sqrt{q\bar{q}}\sim\exp\left(-a\sqrt{kn_{K}\gamma}|c_{s}|/2\right)$ so as $a$ increases the gap narrows
exponentially and vice versa. All these features are supported by the numerically computed stationary solutions of Eq. \eqref{eq:full_vector_NL_eq} (see Fig. \ref{f3}). These eigenstates of $\hat{M}$ are searched in the form $\mathbf{E}(x,z)=\mathbf{E}(x)e^{i\beta z}$ using an iterative Fourier method that fixes $a$, so that different families $P(\mu;a)$ are found separately. Soliplasmons of $\delta=0,\ \pi$ types naturally appear (see insets in Fig. \ref{f3}).

Stability of soliplasmons with frequency $\omega\simeq1.26$ PHz  has been checked by numerical propagations over $z=60\ \mu$m with an input noise of $20\%$ in amplitude at $z=0$ without ohmic losses ($\epsilon_{m}\in\mathrm{Re}$). The stationary states are computed in a single interface configuration between
silica glass, $\epsilon_{K}=\epsilon_{d}=2.09$, $n_{2}=2.6\times10^{-20}\ m^{2}$/W,
and silver $\epsilon_{m}=-82$. Coupling coefficients $q$, $\bar{q}$
are inversely proportional to $N_{p}$, $N_{s}$ and hence a \text{weak} plasmon ($N_s\gg N_p$) is driven by the \textit{soliton} and vice versa. We focus on the former case, $N_{s}\gg N_{p}$: mathematically, $|c_{s}|^{-1}d|c_{s}|/dz\ll|c_{p}|^{-1}d|c_{p}|/dz$ and $\bar{q}\ll q$, so soliton dynamics is essentially quasi-stationary.
The previous assumption together with $c_{p,s}=|c_{p,s}|\exp(i\phi_{p,s})$
allow us to rewrite the Eq. \eqref{eq:variational_eqs} for the SPP amplitude, $|c_{p}|$, and the soliplasmon phase,
$\phi_{sp}\equiv\phi_{p}-\phi_{s}$,

\begin{equation}
\frac{d\phi_{sp}}{dz}\approx2\Delta_{\mu}+q\frac{|c_{s}|}{|c_{p}|}\cos\phi_{sp},\,\,\,\frac{d\left|c_{p}\right|}{dz}\approx q|c_{s}|\sin\phi_{sp}.\label{Ref:eqd4}
\end{equation}

The low nonlinearity of silica glass and the power levels used here
($g\left|c_{s}\right|^{2}\sim10^{-2}$) locate the soliplasmon dispersion
anti-crossing frequency substantially below our pump frequency. For
our particular choice of parameters the stationary solutions satisfy
$2\Delta_{\mu}>q\left|c_{s}\right|/\left|c_{p}\right|>0$, so the right hand side of the phase equation \eqref{Ref:eqd4} is always positive
for the initial conditions ($\beta_{p}$ in the MK system is estimated
from the vertical asymptote in Figs. \ref{f3}).
Eqs. \eqref{Ref:eqd4} tells us that the soliplasmon phase will initially increase regardless the type of soliplasmon and
that $|c_{p}|$ will increase (decrease) for initial 0-($\pi-$) soliplasmons.
These two features are clear in all our propagation simulations, which
integrate Eq.\eqref{eq:full_vector_NL_eq} with no approximation (see
Figs. \ref{f4}-\ref{f6}) and permit to evaluate $c_{p}$ and $c_{s}$
as the peak amplitudes of the plasmon and soliton component of the
complete solution. Figure \ref{f4} shows the propagation of a $\delta=\pi$
solution with $n_{\mathrm{eff}}=1.464$. Apart from being diffraction
free (compare with linear propagation inset), the input noise used
for all the simulations introduces fluctuations which propagate away
from the soliplasmon (see inset). Here the decrease of $|c_{p}|$
is associated to the transfer of energy from the SPP to the soliton, as shown by the stream lines and power flux arrows. 

Propagation
of a $\delta=0$ solution, see Fig. \ref{f5}, shows a very different behavior, since the predicted initial increase
of $|c_{p}|$ implies that the SPP drains energy from the soliton. Remarkably, $d|c_{p}|/dz=0$
provided that $\phi_{sp}=0,\pi$, as Eqs. \eqref{Ref:eqd4} predict
and shown in Fig. \ref{f5}c. At these points we observe that the
exchange of energy is reversed. Qualitatively,
one could explain the dynamics in Fig. \ref{f5} as follows. At $z=0$
the soliton tail reaches the metal interface and pumps the weak SPP
wave. As a result of this power transfer $|c_{p}|$ and $\beta_{p}$
($n_{\mathrm{eff,}p}$) increase whilst $|c_{s}|$ and $\beta_{s}$
($n_{\mathrm{eff},s}$) decrease. This means that the soliton is accelerated
and the SPP is slowed down, varying their relative phase, $\phi_{sp}$. At $z\approx18\ \mu$m, $\phi_{sp}=\pi$
but a $\pi$-soliplasmon with more stable dynamics can not be formed
due to the different velocities between the soliton and SPP ($d\phi_{sp}/dz\neq0$).
As they come back in phase the SPP returns some energy to the
soliton and the initial parameters are approximately restored
at $z=35\ \mu$m. In this particular example, the soliton is slowly
attracted towards the interface, presumably due to the potential
well formed by the SPP in the Kerr medium. Simulations suggest that $\delta=\pi$
soliplasmons are more stable than the $\delta=0$ ones, what can be associated
to the initially smaller value of $d\phi_{sp}/dz$ for our conditions
in the $\delta=\pi$ case. Indeed, Fig. \ref{f5} suggests that SPP excitation from an initial soliton is possible in the $2\Delta_{\mu}>q\left|c_{s}\right|/\left|c_{p}\right|>0$ regime, because the soliton tail at the metal interface has zero relative phase with the soliton core, what results in a periodic transfer of energy between the soliton and SPP.

For $2\Delta_{\mu}<q\left|c_{s}\right|/\left|c_{p}\right|<0$ ($\omega\ll\omega_m$), $\pi$-soliplasmons are expected to undergo a more rapidly varying dynamics than $\delta=0$ ones. This regime requires higher $\varepsilon_K$ in the MDK configuration and will be reported elsewhere. Stability here is defined from the dynamics associated to $|c_{p,s}|$ and $\phi_{sp}$ along $z$, rather than to the \textit{rupture} of the nonlinear states, which is not possible since both solitons and SPP's are stable solution separately.

Effect of ohmic losses is shown explicitly in Fig. \ref{f6}, which has identical initial conditions as in Fig. \ref{f5}.
The first stages of the dynamics are very similar in the two cases.
However, now the soliton acts as a reservoir for the plasmon (note
that $|c_p(30\ \mu m)|\approx|c_p(0\ \mu m)|)$. As soon as the phase reaches $\phi_{sp}=\pi$,
the energy transfer is frustrated and $|c_{p}|$ drops dramatically due to losses. This reduces $\beta_{p}$ and $\phi_{sp}$
decreases towards the initial value $\phi_{sp}$. Exposure to metal
reflection bends the soliton trajectory and it goes away from the
interface, leaving behind a SPP that will be exponentially attenuated.

In summary, we have analyzed the stationary and dynamical properties of soliton-plasmon bound states, \textit{soliplasmons}, by means of a simplified variational model, obtained from Maxwell equations. This model predicts the relevant physics associated to these hybrid nonlinear waves and is in good agreement with first principle modeling.  Our novel model opens the possibility to study new power-tunable photonic devices based on nonlinear soliplasmonic waveguides. This work was partially supported by Contract No. TEC2010-15327.


\end{document}